\documentclass{WileyMSP-template}

\usepackage{graphicx}
\usepackage{fancyhdr}
\graphicspath{{./Figures/}}

\usepackage{bm}

\usepackage{amsmath}
\usepackage{siunitx}
\newcommand{\chlamy}[1][{ }]{\textit{Chlamydomonas reinhardtii#1}} 

\usepackage{booktabs}

\usepackage{xr}
\usepackage{xcolor}

\usepackage{ulem}

\usepackage[super,square,sort&compress,comma]{natbib}

\begin{document}

\pagestyle{fancy}
\rhead{\includegraphics[width=2.5cm]{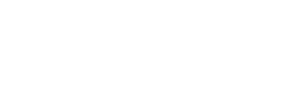}}

\title{Permeation dynamics of active swimmers through anisotropic porous walls}

\maketitle


\author{Florian von R\"{u}ling*}
\author{Liubov Bakhchova}
\author{Ulrike Steinmann}
\author{Alexey Eremin}


\dedication{}

\begin{affiliations}
F.  von R\"{u}ling,  Prof.  Dr.  A. Eremin\\
Department of Nonlinear Phenomena\\
Otto von Guericke University Magdeburg\\
Universit\"{a}tsplatz 2, 39106 Magdeburg,  Germany\\
Email Address: florian.vonrueling@ovgu.de

L.  Bakhchova,  Prof.  Dr.  U.  Steinmann\\
Institute for Automation Engineering\\
Otto von Guericke University Magdeburg\\
Universit\"{a}tsplatz 2, 39106 Magdeburg,  Germany

\end{affiliations}


\begin{abstract}

Natural habitats of most living microorganisms are distinguished by a complex structure often formed by a porous medium such as soil. The dynamics and transport properties of motile microorganisms are strongly affected by crowded and locally anisotropic environments.
Using \chlamy as a model system, we explore the permeation of active colloids through a structured wall of obstacles by tracking microswimmers' trajectories and analysing their statistical properties. Employing micro-labyrinths formed by cylindrical or elongated pillars, we demonstrate that the anisotropy of the pillar's form and orientation strongly affects the microswimmers' dynamics on different time scales. Furthermore, we discuss the kinetics of the microswimmer exchange between two compartments separated by an array of pillars.
\end{abstract}

\section{\label{sec:intro}Introduction}
In laboratories, microswimmers are often studied either in bulk~\cite{Drescher2010,Howse2007,Arrieta2017,Polin2009} or,  facilitating the observation,  in homogeneous quasi-two-dimensional geometries such as thin films~\cite{Guasto2010}, narrow Hele-Shaw-cells~\cite{Bregulla2014} or confined to interfaces~\cite{Wang2015}. In contrast,  the natural habitats of swimming microorganisms are very complex. Biological self-propelled particles like \chlamy inhabiting soil~\cite{Sasso2018} or \textit{E. coli} dwelling in mammalian guts~\cite{Blount2015} encounter non-planar, rough surfaces and interact with suspended and sedimented passive particles. Through constraints, such a complex environment profoundly affects the dynamics of the microswimmers. 

It is essential to understand how microswimmers steer through complex environments to comprehend phenomena like groundwater contamination~\cite{Haznedaroglu2010} and the spreading of infections in animals~\cite{Sultan2013} or plants~\cite{Hossain2005}.
Potential applications of self-propelled particles,  e.g.  bacterial bioremediation~\cite{Singh2008} or tumor treatment~\cite{Duong2019}, would benefit from a profound knowledge about the effects of  heterogeneous media on microswimmer motility.
 
 The mechanisms underlying the active scattering of \chlamy have been explored in a series of experimental and simulation studies~\cite{Kantsler2013,Contino2015,Lushi2017,Mirzakhanloo2018}. Since the scattering of active swimmers at obstacles is determined by complex hydrodynamic and steric interactions and involves the nonequilibrium phenomenon of self-propulsion,  it differs substantially from specular reflection. Experimental investigation of the interactions of the motile puller-type algae with a planar wall~\cite{Kantsler2013} revealed a constant outgoing angle determined by the swimmer geometry. Interactions with cylindrical obstacles are more complex and can be purely hydrodynamic~\cite{Contino2015}. In arrays of large axisymmetric pillars,  preferred swimming directions arise in an ensemble of \chlamy~\cite{BrunCosmeBruny2019}. 
 
\begin{figure*}
\centering
\includegraphics[width=\textwidth]{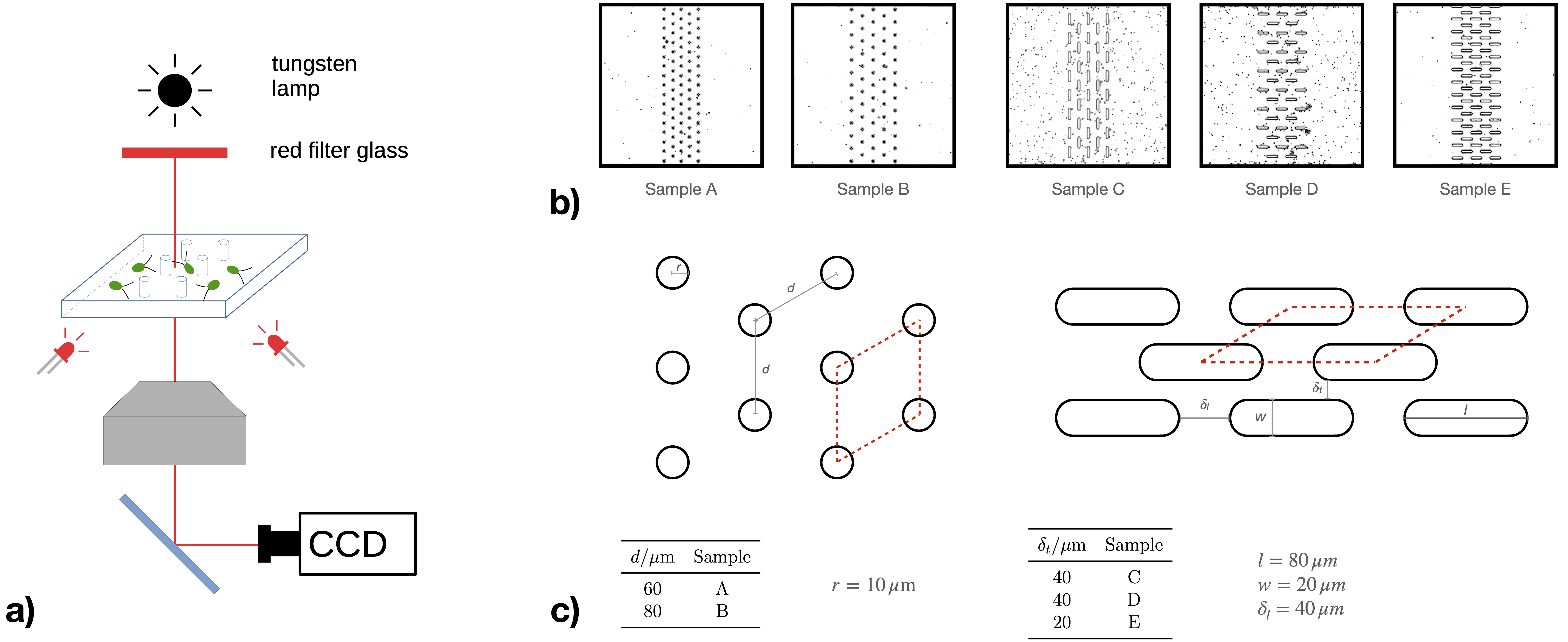}
\caption{a) Sketch of the setup employed for the observation of \chlamy in patterned microchannels. b) Microchannel samples \textbf{A} - \textbf{E} consisting of two chambers separated by porous walls. c) Illustration of the geometry of pillar lattices in the microchannels and differentiation of the samples.  Samples \textbf{A}, \textbf{B} contain cylindrical pillars and differ in the center-to-center-distance between adjacent pillars \(d\).  The pillars in samples \textbf{C},\textbf{D},\textbf{E} are stadium-shaped pillars and the lattices vary with respect to the pillar orientation and the transversal inter-pillar-distance \(\delta_t\). }
\label{fig:Samples}
\end{figure*}

 Thin microchannels with a thickness of \qty{20}{\micro\meter}, slightly larger than the cell diameter, restrict the motion of the swimming algae to a single plane. Such quasi-2D experimental geometry considerably simplifies the analysis of microswimmers' trajectories and allows exploring the confinement effects. In this paper, we explore the exchange of microswimmers \chlamy between two compartments separated by a porous wall modelled by an array of pillars. Choosing rounded and elongated pillars (Figure \ref{fig:Samples}) allows us to study the effect of the anisotropy on the exchange kinetics and the character of microswimmers' trajectories. We demonstrate that the pillar's form and orientation strongly affect the transmission and reflection coefficients of the swimming algae and the orientational anisotropy of their motion.

\section{\label{sec:results}Results and discussion }
\subsection{Equilibration kinetics}
In the microchannels,  an array of pillars serving as a model system for a porous wall separates two obstacle-free compartments \textbf{1} and \textbf{2}.  
Excursions of a microswimmer through the porous wall may either result in the swimmer's permeation from one compartment  to the other ("transmission") or its reflection,  i.e. the swimmer's return to the obstacle-free region it came from initially. A distribution mismatch in microswimmers can be induced by localised illumination. In thin microchannels ($\approx \qty{20}{\micro\meter}$ thickness), we could not observe any noticeable phototaxis; however, illumination by blue light (\(\qty{480}{nm}\)) induced adsorption of the microswimmers at the glass substrate.  

\begin{figure}[ht!]
   \centering
    \includegraphics[width=0.5 \linewidth]{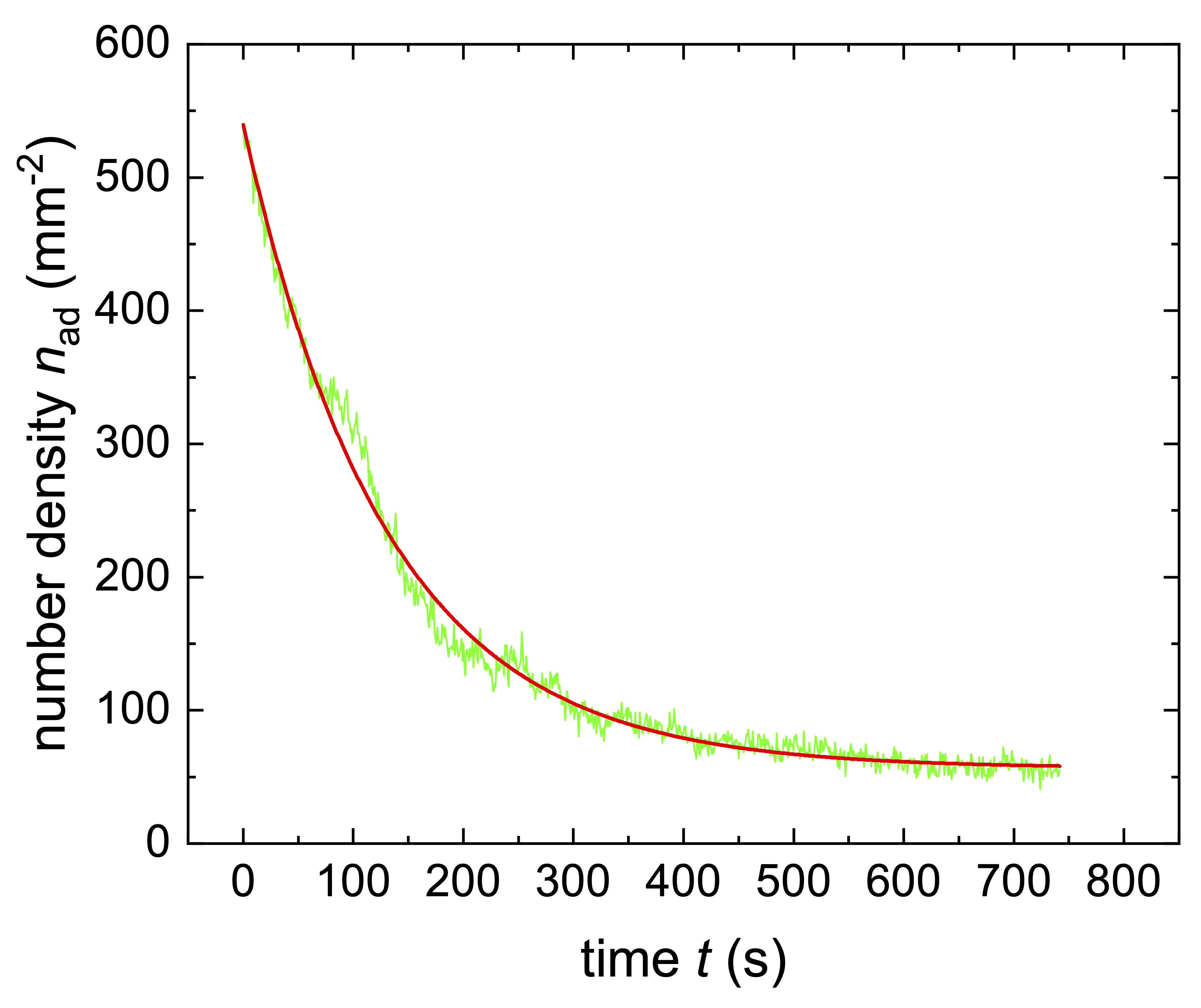}
    \caption{Desorption kinetics: Time dependence of the number density of adhered (non-swimming) cells in the compartment \textbf{1} which had been illuminated by localised blue light before the measurement. The red line is the exponential fit.} 
    \label{fig:adhered_n}
\end{figure}

\begin{figure}[ht!]
\centering
   \includegraphics[width=0.5 \linewidth]{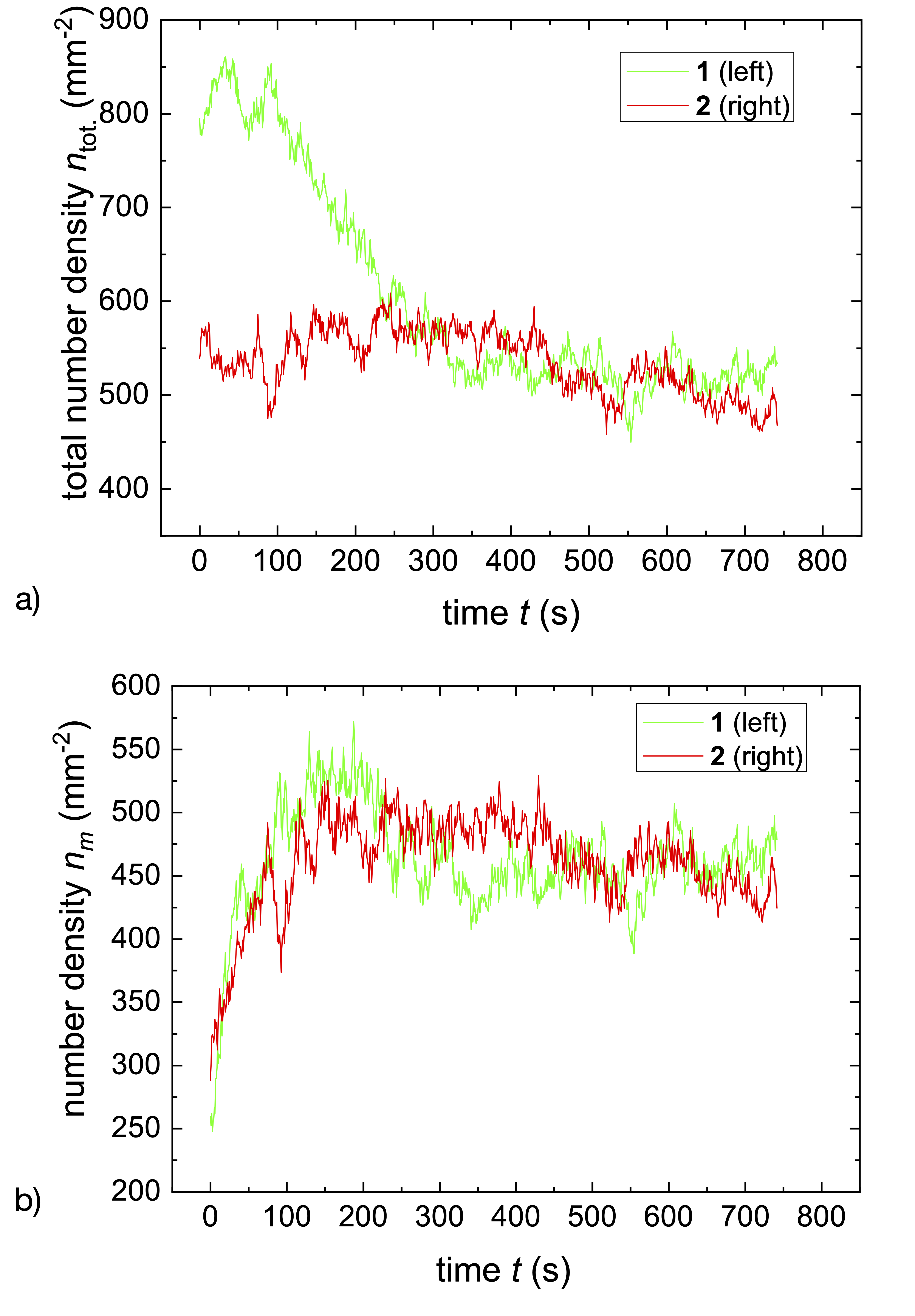}
    \caption{Equilibration of the microswimmers between two compartments after termination of the local illumination in compartment \textbf{1}. (a) Total number density of motile and immotile (adhered) swimmers, (b) number density of motile swimmers only. } 
    \label{fig:equillibration}
\end{figure}

Through adhesion, immotile algae were collected in the chamber exposed to the local illumination. We designate the algae in the adhered state as 'immotile' although they retain slow gliding motility \cite{Bloodgood1981, Till2022}. When the local illumination was terminated, the cells desorbed and transitioned into the motile state (Figure~\ref{fig:adhered_n}). Figure~\ref{fig:equillibration} shows the equalisation of the microswimmer number density in two compartments after the illumination was removed.  The numbers equalise on a time scale of about \qty{200}{\second}.  Although there is a light-induced disbalance in the total number density of  cells \(n_\text{tot}=n_\text{m}+n_\text{ad}\) which slowly equilibrates (Figure~\ref{fig:equillibration}a), the number density of swimming algae \(n_\text{m}\) equilibrates between the two compartments nearly instantaneously (Figure~\ref{fig:equillibration}b). This can be attributed to the low rate of the desorption process (Figure~\ref{fig:adhered_n}). During detachment, the number of adhered cells exhibits a nearly exponential decay to the number determined by the global illumination intensity and other conditions of the cells. 

 The characteristic desorption time $\tau_{\mathrm{dsp}}$ can be extracted from the single exponential fit 
\begin{equation}
 n_{\mathrm{ad}}(t)=n_{\mathrm{ad}}^{\infty}\exp{(-t/\tau_{\mathrm{dsp}})}
  \end{equation} 
  giving \(\tau_{\mathrm{dsp}}=\qty{135}{\second}\).

 \subsection{Near wall accumulation}
 
\begin{figure}
\centering
\includegraphics[width=0.5 \linewidth]{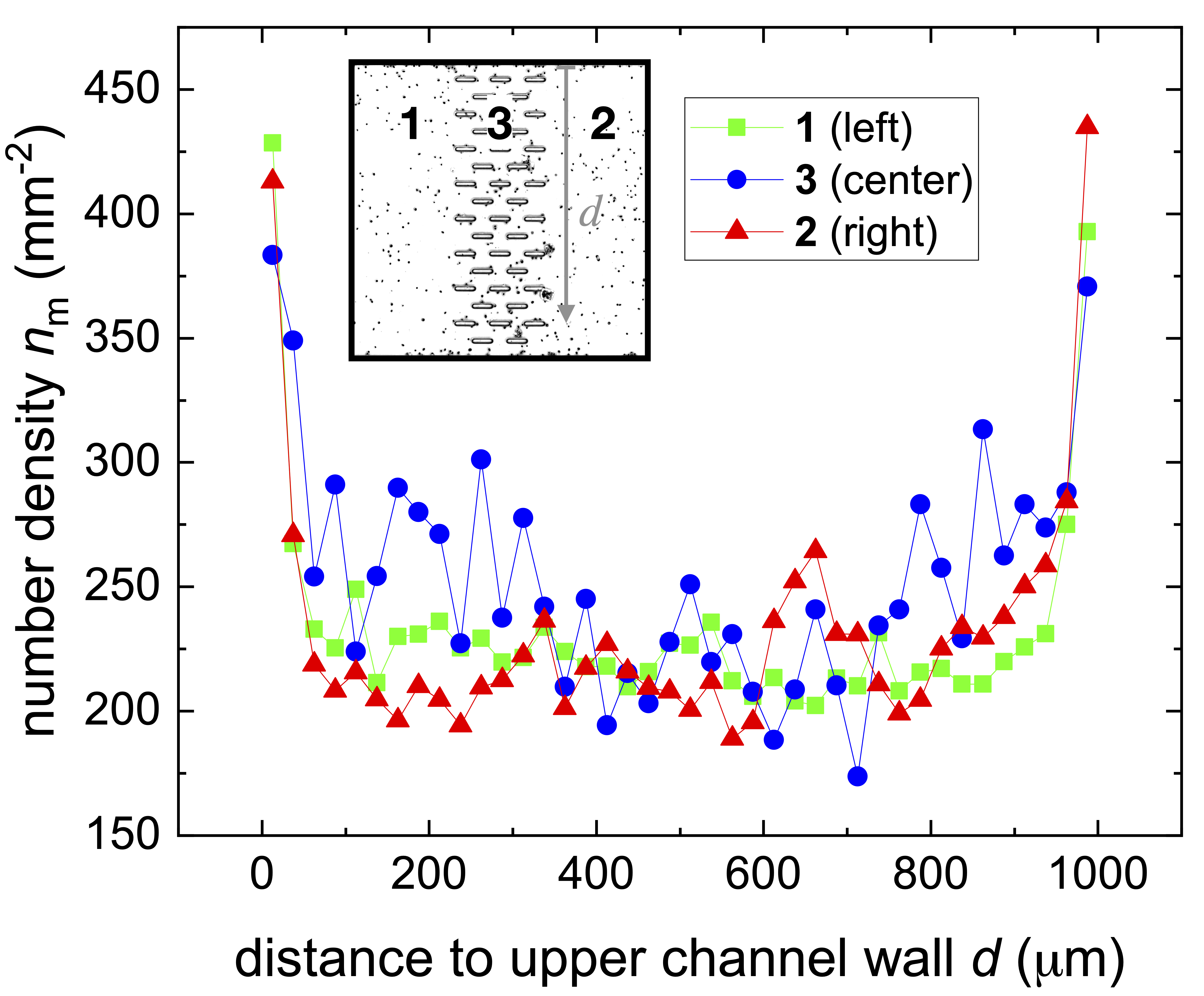}
\caption{The time-averaged cell number density \(n_{\text{m}}\) depending on the distance from the upper microchannel wall \(d\) for Sample \textbf{D}.  Compartment \textbf{3} (blue curve) contains pillars parallel to the channel walls.  The inset illustrates the definition of \(d\) and the division of the channel into three compartments.  }
\label{fig:accumulation}
\end{figure}

In contrast to the quick equilibration of the number of moving cells  between compartment \textbf{1} and \textbf{2},  there is a disbalance of the cell concentration across the width of the microchannel as the swimmers accumulate at the channel walls.  This behaviour has been reported for \chlamy by Williams \textit{et al.} \cite{Williams2022}.  Surprisingly,  the accumulation at the microchannel walls seems to be largely unaffected by the microlabyrinths even for Sample \textbf{D} with elongated pillars oriented parallel to the channel walls. (see Figure \ref{fig:accumulation}).

\subsection{Cylindrical obstacles}

\begin{figure}[ht!]
\centering
    \includegraphics[width=0.5 \linewidth]{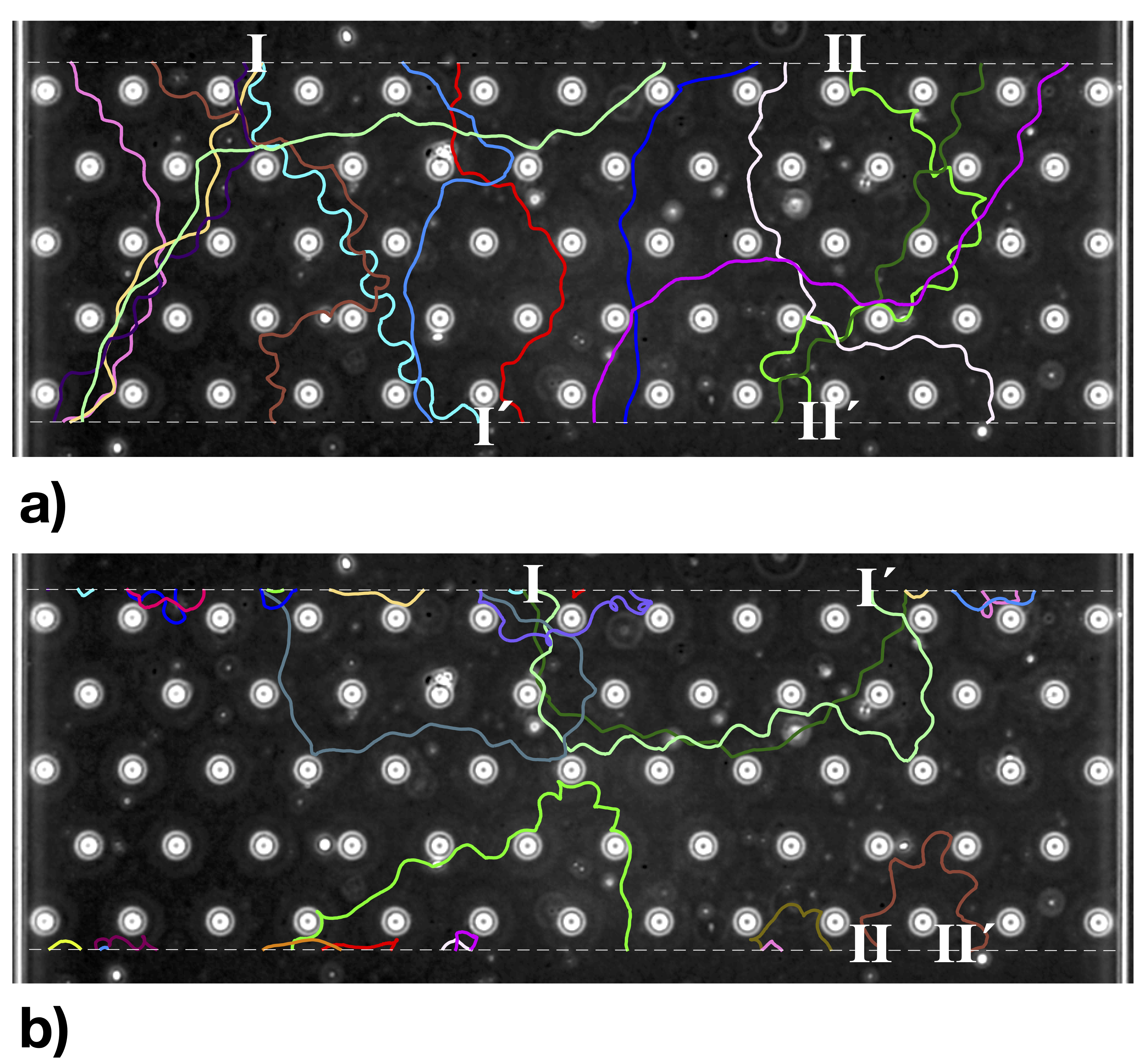}
    \caption{Selected trajectories in the compartment \textbf{3}  of Sample \textbf{D} containing cylindrical pillars. a) Trajectories resulting in the microswimmers' permeation from one compartment to another. b) Trajectories of microswimmers which start and end in the same compartments (reflections). Starting and ending points for two selected trajectories are designated by the pairs I-\(\text{I}^\prime\) and II-\(\text{II}^\prime\).  The boundaries of compartment \textbf{3} (dashed lines) are located \(\qty{17}{\mu m}\) from the outermost obstacle row, which represents the distance below which steric swimmer-obstacle interactions can occur (assuming a swimmer body radius of \(\qty{5}{\mu m}\) and a flagellar length of \(\qty{12}{\mu m}\) \cite{Jeanneret2016}).  } 
    \label{fig:trajectories_cyl}
\end{figure}

The effect of geometrical constraints on the swimming behaviour is manifested in the character of the trajectories. Scattering of the algae at the obstacles results in the reduction of the swimming persistence. At the same time, regularly ordered obstacles affect the alignment of the microswimmers' trajectories achieved through sliding over the surface of the pillars. Typical trajectories for the Sample \textbf{A} in the compartment \textbf{3} containing the array of pillars are shown in Figure~\ref{fig:trajectories_cyl}.

Most trajectories exhibit small undulations.
\chlamy are known to swim along helical trajectories in bulk \cite{Polin2009},  which can be attributed to the nonplanar flagellar beating and a small asymmetry in the flagellar driving forces \cite{Cortese2021}.
The capillary confinement restricts the rotation of the algae and suppresses their helical motion. Instead, undulated character of the trajectories occurs.  A combination of hydrodynamic and steric interactions
 of the algae with the pillars results in their scattering, which has been studied by Contino \textit{et al.}\cite{Contino2015}.  In a periodic array of pillars,  preferred swimming directions along lanes between the pillars arise \cite{BrunCosmeBruny2019}.
 
  The trajectories can start in one compartment and end in the other compartment, as exemplarily shown in Figure~\ref{fig:trajectories_cyl}a. This case corresponds the permeation through compartment \textbf{3} to get from \textbf{1} to \textbf{2} or from \textbf{2} to \textbf{1}. Some algae can even turn around and, after an excursion through the compartment \textbf{3}, return to the same compartment they came from (Figure  \ref{fig:trajectories_cyl}b). This situation is designated as the reflection.

\begin{figure}[ht!]

   \centering
   \includegraphics[width=0.5 \linewidth]{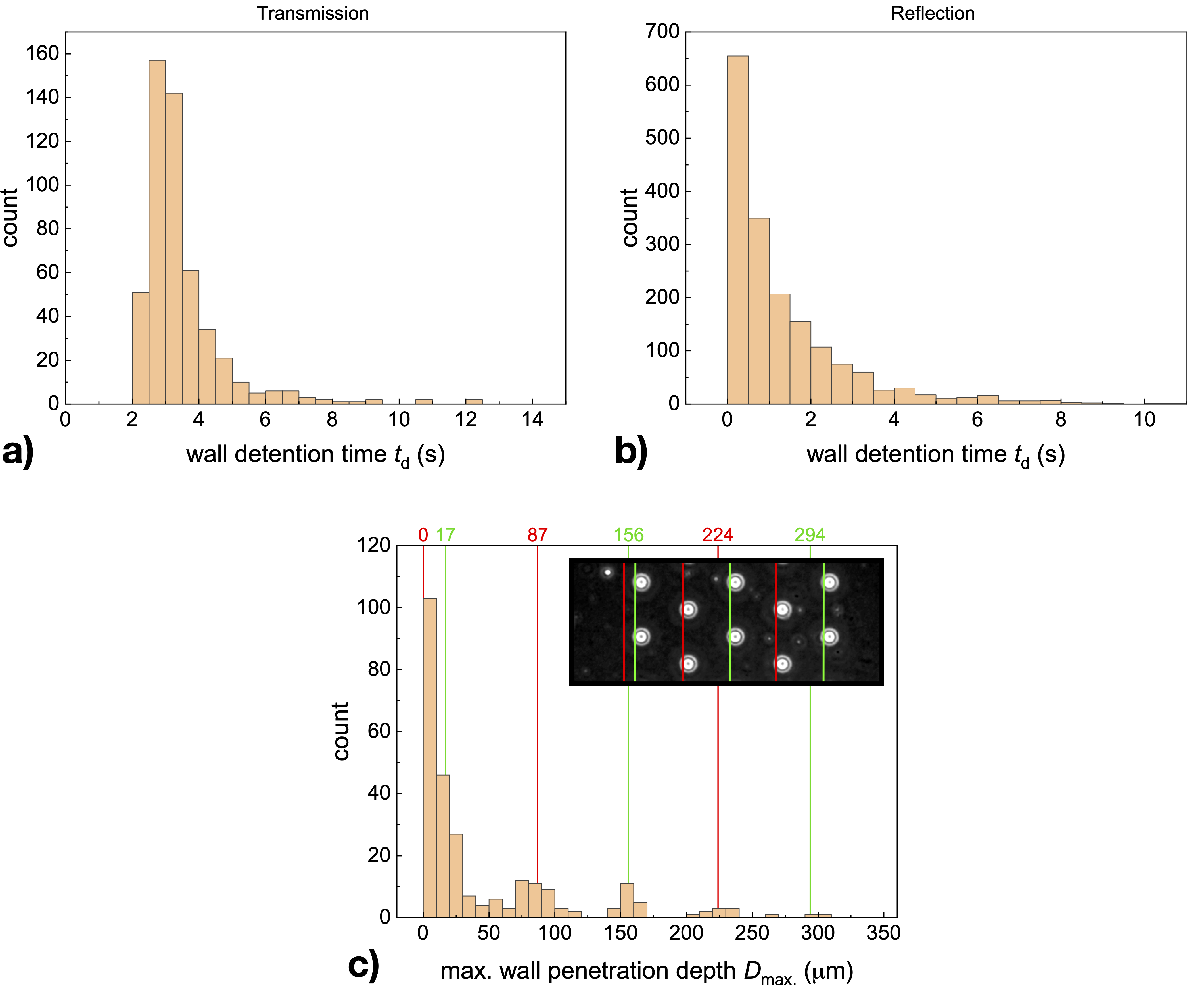}
    \caption{Distribution of wall detention times for a) permeating and b) reflecting trajectories for Sample \textbf{B} with cylindrical pillars. c) Corresponding distribution of the maximal penetration depth.} 
    \label{fig:cyl_trajectories_stat}
\end{figure}

Although any alga can permeate or reflect, the reflection and permeation events strongly differ in their statistics (Figure~\ref{fig:cyl_trajectories_stat}). Reflecting algae have significantly smaller mean detention times in compartment \textbf{3} than permeating ones ($\tau_{\mathrm{det}}^{\mathrm{ref}}\approx \qty{2.8}{\second}$, $\tau_{\mathrm{det}}^{\mathrm{perm}}\approx \qty{8.5}{\second}$ for Sample \textbf{B}). This can be explained by the fact that most algae scatter back at the first row of pillars. This can be well seen in Figure~\ref{fig:cyl_trajectories_stat}, where the distribution of the maximal penetration depth for reflecting trajectories is shown. The distribution function exhibits multiple maxima corresponding to the positions of the pillar planes. This suggests that the scattering effects are responsible for the reflection rather than single cells' tumbling or meandering motion (see also Figure S2).

\begin{figure}
\centering
\includegraphics[width=0.5 \linewidth]{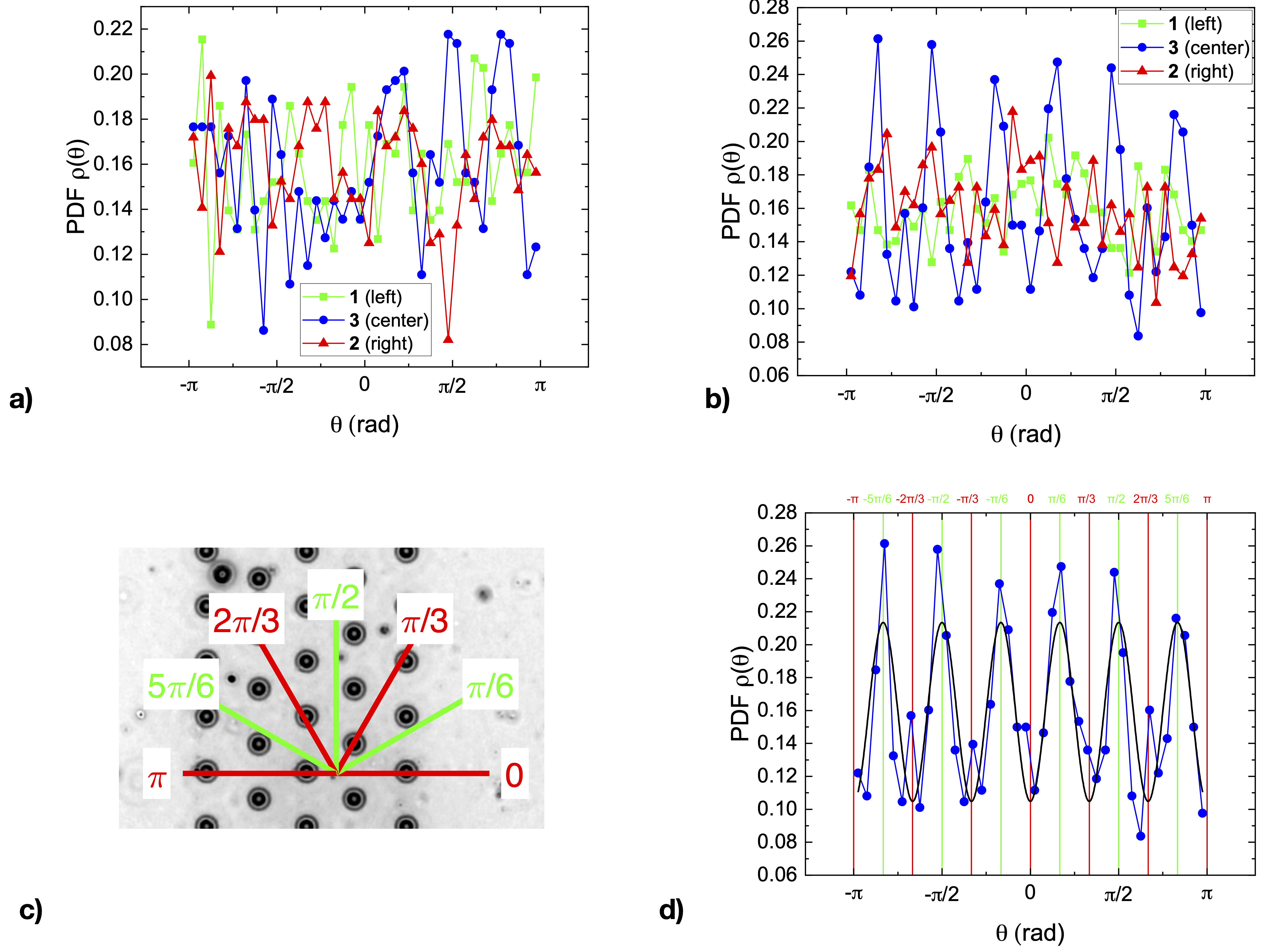}
\caption{Angular probability distribution functions of displacements over \(\Delta t =\qty{2}{s}\) in a) Sample \textbf{B} and b) Sample \textbf{A} in different compartments.  Distinguished directions in the periodic pillar array are shown in c).  d) shows the PDF for the compartment \textbf{3} in Sample \textbf{A} with the fit according to Equation \eqref{eq:FitHex} (black line).}
\label{fig:angularPDFCyl}
\end{figure}

Preferred directions of motion in the lattice of pillars are revealed in orientational probability distribution functions (PDF) of displacements.
In the case with high porosity (Sample \textbf{B},  see Figure \ref{fig:angularPDFCyl}a),  the PDF shows isotropic distribution.  Anisotropy arises as the inter-pillar-distance is decreased (Sample \textbf{A},  see Figure  \ref{fig:angularPDFCyl}b).  Peaks in the angular PDF  correspond to the directions along which free straight paths exist (Figure \ref{fig:angularPDFCyl}c,d).  Motion along the directions for which dips in the PDF occur would result in collisions with obstacles leading to a reorientation of the algae. 

Swimming anisotropy has previously been observed by Brun-Cosme-Bruny \textit{et al.} \cite{BrunCosmeBruny2019} in a square lattice of large cylindrical pillars.  In that case,  a smaller number of distinguished directions emerged,  apparently due to the difference in obstacle size \cite{Reichhardt2020a} and lattice geometry. Following  \cite{BrunCosmeBruny2019}, the swimming anisotropy can be analytically accounted for by introducing an anisotropic scattering rate. Without obstacles, \chlamy are modelled as active Brownian particles exhibiting rotational diffusion characterised by the constant \(D_r\). 

 In the presence of obstacles, the swimmers experience additional random direction changes due to collision-induced tumbling.  The tumbling rate \(\lambda\) depends on the direction of motion. In the case of a square lattice,  described in Ref. \cite{BrunCosmeBruny2019},  \(\lambda (\theta) = \lambda_0 - \lambda_4 \cos(4\theta)\) and \(\lambda_0=\lambda_4\) such that the tumbling rate vanishes in directions \(0,\pm\pi/2,\pi\) along which obstacle-free lanes exist.  To adapt the model to the case of a hexagonal lattice of cylindrical obstacles,  we describe the tumbling anisotropy via 
\begin{equation}
\lambda (\theta) = \lambda_0 + \lambda_6 \cos(6\theta)
\end{equation}
  The angular distribution function,  following the derivation described in Ref.  \cite{BrunCosmeBruny2019},  is 
\begin{equation}
\bar{P} (\theta) = \frac{1}{2\pi}\left(1-\frac{\lambda_6}{36 D_r+\lambda _0} \cos (6\theta)\right)
\label{eq:FitHex}
\end{equation}
The experimental data can be fitted with Equation \eqref{eq:FitHex} assuming \(\lambda_6=\lambda_0\) (see Figure  \ref{fig:angularPDFCyl}d),  showing qualitative agreement with the description as an anisotropic scattering medium.

\subsection{Shape-anisotropic obstacles}
Shape-anisotropic obstacles, in the form of elongated pillars, strongly affect the dynamics of microswimmers in the compartment \textbf{3} exhibiting a different behaviour than that observed in the case of cylindrical pillars. Pillars' shape and orientation also affect the exchange rates between the compartments \textbf{1} and \textbf{2}. 
\begin{figure*}[ht!]
\centering
\includegraphics[width=0.5\linewidth]{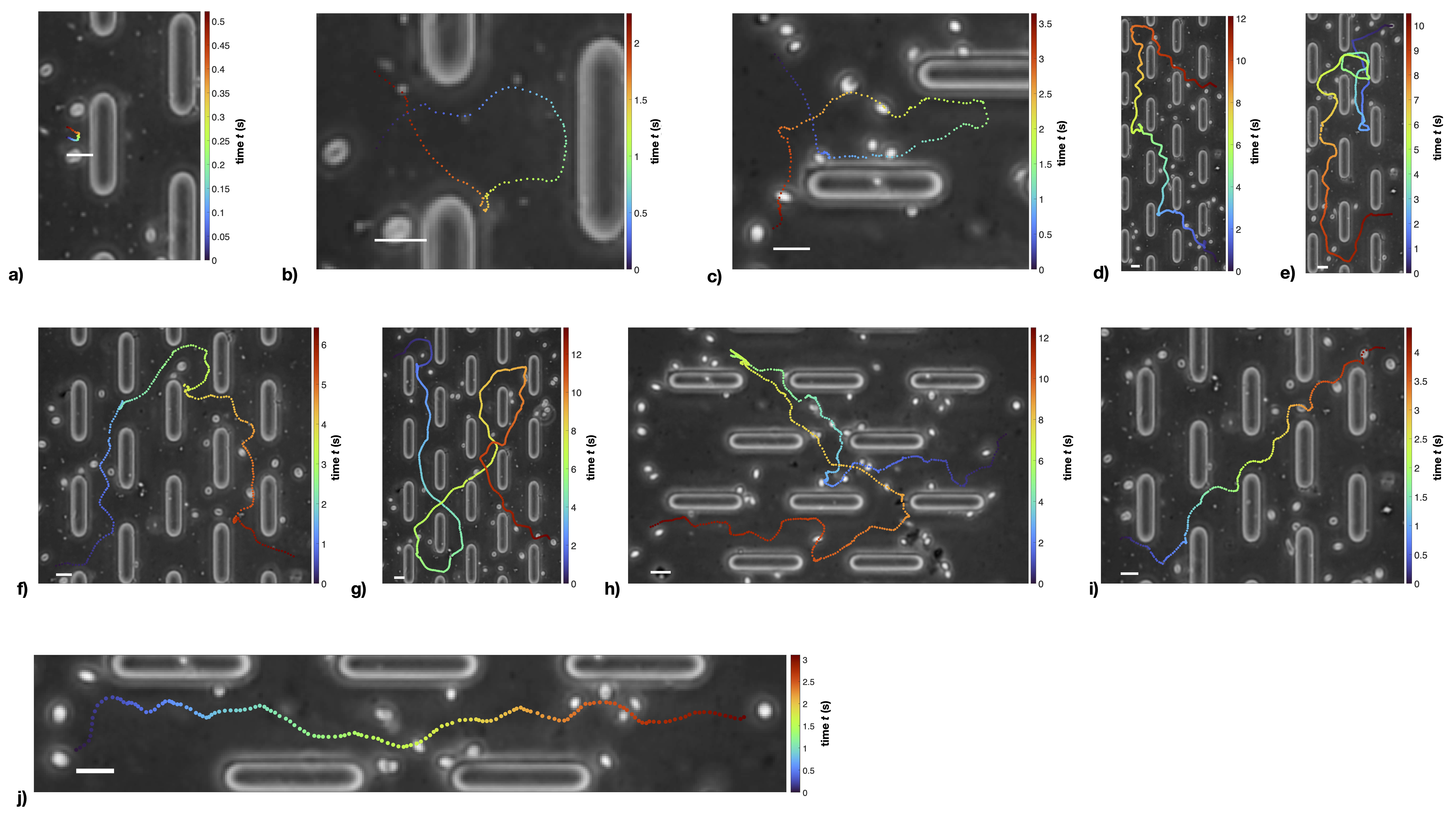}
\caption{Exemplary trajectories of \chlamy in the presence of elongated obstacles.  The background image is a phase contrast micrograph of the microchannel containing the motile algae. The long obstacle axes was either perpendicular (vertical) or parallel (horizontal) to the microchannel walls.  Scalebars have a length of \(\qty{20}{\micro\meter}\). }
\label{fig:Trajectories}
\end{figure*}
Exemplary trajectories of the motile algae in the porous environment of compartment \textbf{3} are shown in Figure  \ref{fig:Trajectories} for elongated obstacles with their long axes perpendicular or parallel to the microchannel walls.   

Similarly to the case of cylindrical pillars, reflections of the algae often occur at the outermost obstacle row (Figure  \ref{fig:Trajectories}a).  This is especially well seen for orthogonal pillar orientation.  Note that yet,  trajectories of reflection events can exhibit more complexity (Figure  \ref{fig:Trajectories}b-e). The microswimmers can penetrate deeper into the obstacle array (Figure  \ref{fig:Trajectories}d,e).  
Algae that fully traverse the porous environment may find a straight path (Figure  \ref{fig:Trajectories}i,j) or meander through the labyrinth of pillars (Figure  \ref{fig:Trajectories}g,h).  
Hence,  there is a wide spread of transmission path lengths and of the time spent in the obstacle array until the other compartment is reached.  The trajectory in Figure \ref{fig:Trajectories}h exhibits transient adhesion for less than 2 s.   Owing to frequent encounters with the upper and lower microchannel wall in strong vertical confinement,  this phenomenon,  as well as periods of slow translation superimposed with a 'jiggle motion' are sometimes observed.

\begin{figure*}
\centering
\includegraphics[width=\linewidth]{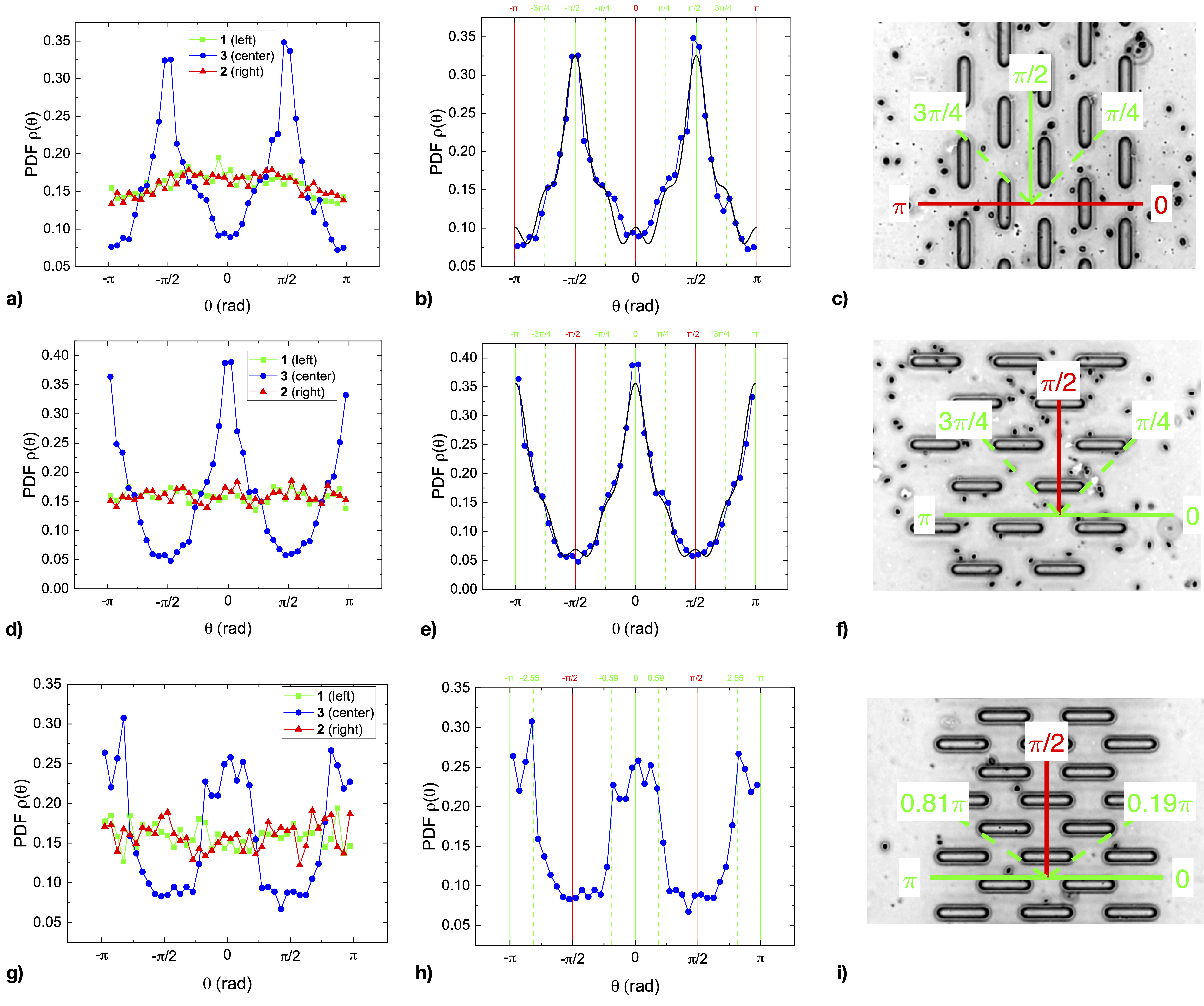}
\caption{Angular PDF of the swimmer displacement over time periods of \(\Delta t = \qty{2}{s}\).  Each row of figures corresponds to one of the Samples \textbf{C} (a-c),  \textbf{D} (d-f) and \textbf{E} (g-i).  The first column (a,d,g) shows the PDF in different compartments.  Figures in the second column (b,e,h) show only the PDF in the compartment \textbf{3} containing pillars.  Vertical lines mark distinguished directions.  For Samples \textbf{C} (b) and \textbf{D} (e) black curves show the fit obtained by applying the effective anisotropic scattering medium model introduced in Ref.  \cite{BrunCosmeBruny2019} to the direction-dependent tumbling rate Equation  \eqref{eq:TumblingRateElongated}.  The third column of figures (c,f,i) illustrates the distinguished directions in the different geometries. }
\label{fig:DisplacementElongated}
\end{figure*}

The environment's anisotropy profoundly affects the orientational order of microswimmers' displacements inside the walls of pillars. 
In the case of the elongated pillars, the PDE exhibit distinctive maxima in the directions parallel to the pillars' long axes (Figure~\ref{fig:DisplacementElongated}).  This confirms the orientational ordering effect determined by the geometrical constraints. Shoulders in the orientational PDE correspond to the specific directions along which free straight paths exist,  \(\pm \pi/4\) and \(\pm 3\pi/4\), for Samples \textbf{C} and \textbf{D} (Figure \ref{fig:DisplacementElongated}b,c and e,f).  Similarly,  peaks in the PDF at approximately \(\pm  0.19\pi,\pm 0.81 \pi\) correspond to obstacle-free directions in Sample \textbf{E} (Figure \ref{fig:DisplacementElongated}h,i). The effective anisotropic scattering medium model proposed by Brun-Cosme-Bruny \textit{et al.} \cite{BrunCosmeBruny2019} can be extended for application to Samples \textbf{C} and \textbf{D}.  The direction dependence of the tumbling rate is modelled as
\begin{equation}
\lambda (\theta) = \lambda_0 - (\lambda_2 \cos(2\theta) + \lambda_4 \cos(4\theta) + \lambda_8 \cos(8\theta))
\label{eq:TumblingRateElongated}
\end{equation}
where \(\lambda_0 \geq 0 \) and \(\lambda (\theta) \geq 0\,\forall \theta \in [-\pi,\pi)\).
The PDF $\bar{P}(\theta)$ has a similar form as Equation  \eqref{eq:FitHex} but containing a sum of cosine-functions \(\cos(k\theta)\) where \(k\in \{2,4,6,8\}\) and with different coefficients.  One should note that the PDF,  other than the tumbling rate,  contains a \(\cos (6\theta)\)-term.  While the shape of the PDF is in good agreement with the experimental data (Figure \ref{fig:DisplacementElongated}b,e),  the tumbling rate is orders of magnitude larger than expected.  There are several possible reasons for the deviation. Firstly,  Samples \textbf{C} and \textbf{D} have a high cell concentration,  where reorientations may be caused by cell-cell-interactions.  The model,  however,  is derived for a single cell.  Secondly,   orientation changes by collision-induced tumbling are modelled by choosing a random angle from the uniform distribution.  Yet,  elongated obstacles,  similar to planar walls \cite{Kantsler2013},  might cause systematic scattering. Indeed,  interactions with the long sides of the elongated obstacles appears to be biased towards small outgoing angles (see Figure S5).  The existence of preferred scattering angles could also explain,  why not all directions with obstacle-free lanes are equal,  i.e.  why directions of motion along the long axis of the obstacles appear as strong peaks in the angular PDF,  whereas the free paths along \(\pm \pi/4\) and \(\pm 3\pi/4\) result in less pronounced shoulders. Thirdly,  elongated obstacles may not only cause systematic scattering but they can even suppress tumbling and promote swimmer alignment.

Despite the alignment induced by the anisotropy of the surrounding,  typical trajectories retain their undulatory form (see e.g.  Figure  \ref{fig:Trajectories}f).

\begin{figure}[ht!]
\centering
\includegraphics[width=0.5\linewidth]{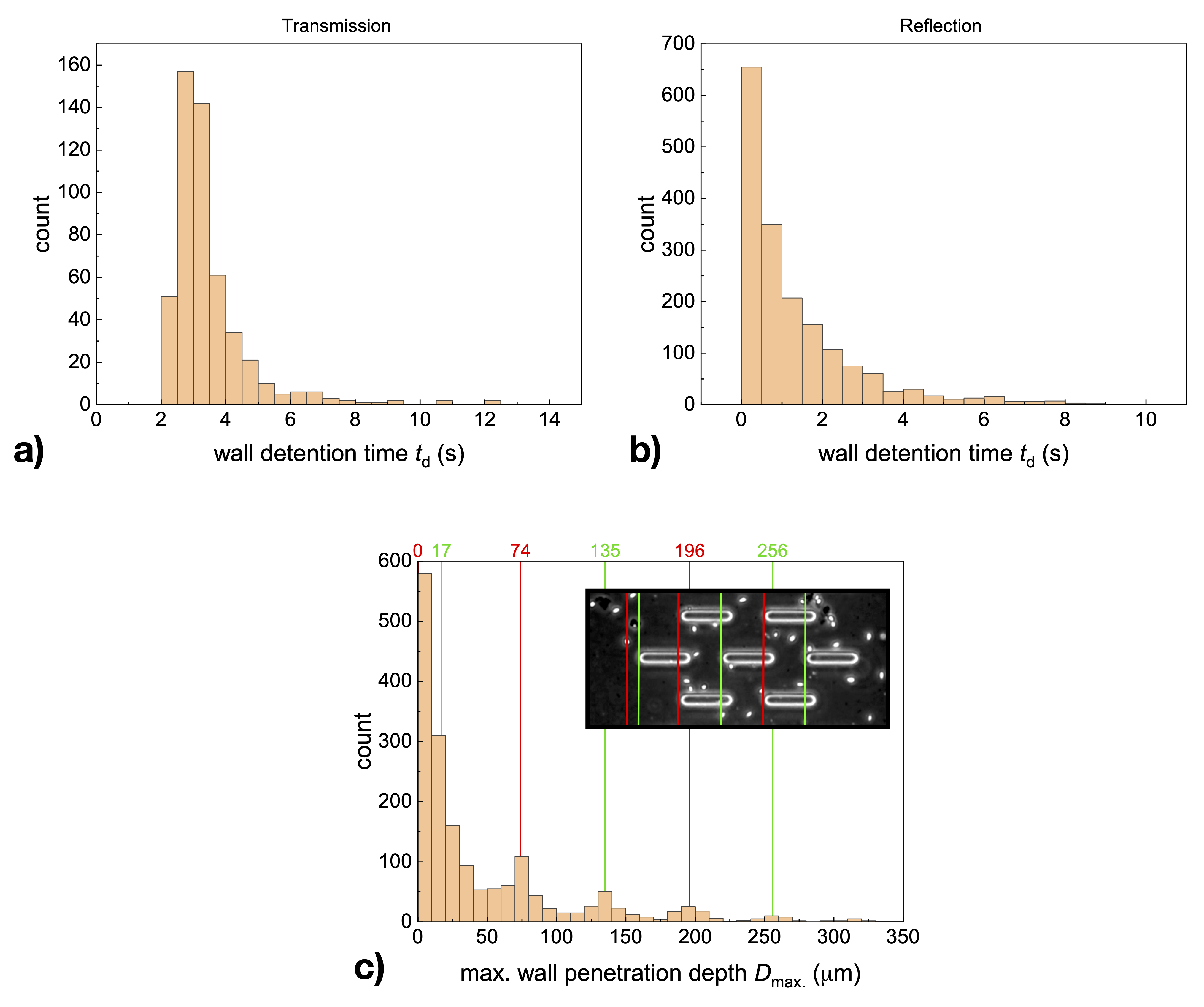}
\caption{Distribution of wall detention times for a) permeating and b) reflecting trajectories for Sample \textbf{D} with elongated pillars. c) Corresponding distribution of the maximal penetration depth. }
\label{fig:par_trajectories_stat}
\end{figure}

For parallel elongated obstacles (Sample \textbf{D}), like in the case of cylindrical pillars, reflecting algae have significantly smaller mean detention times $\tau_{\mathrm{det}}^{\mathrm{ref}}\approx \qty{1.3}{\second}$ in compartment \textbf{3} then permeating ones ($\tau_{\mathrm{det}}^{\mathrm{perm}}\approx \qty{3.5}{\second}$) (Figure~\ref{fig:par_trajectories_stat}a,b). The reflections occur at the tips of the pillars which correspond to the maxima in the penetration length distribution function (Figure~\ref{fig:par_trajectories_stat}c).

\begin{figure}[ht!]
\centering
\includegraphics[width=0.5 \linewidth]{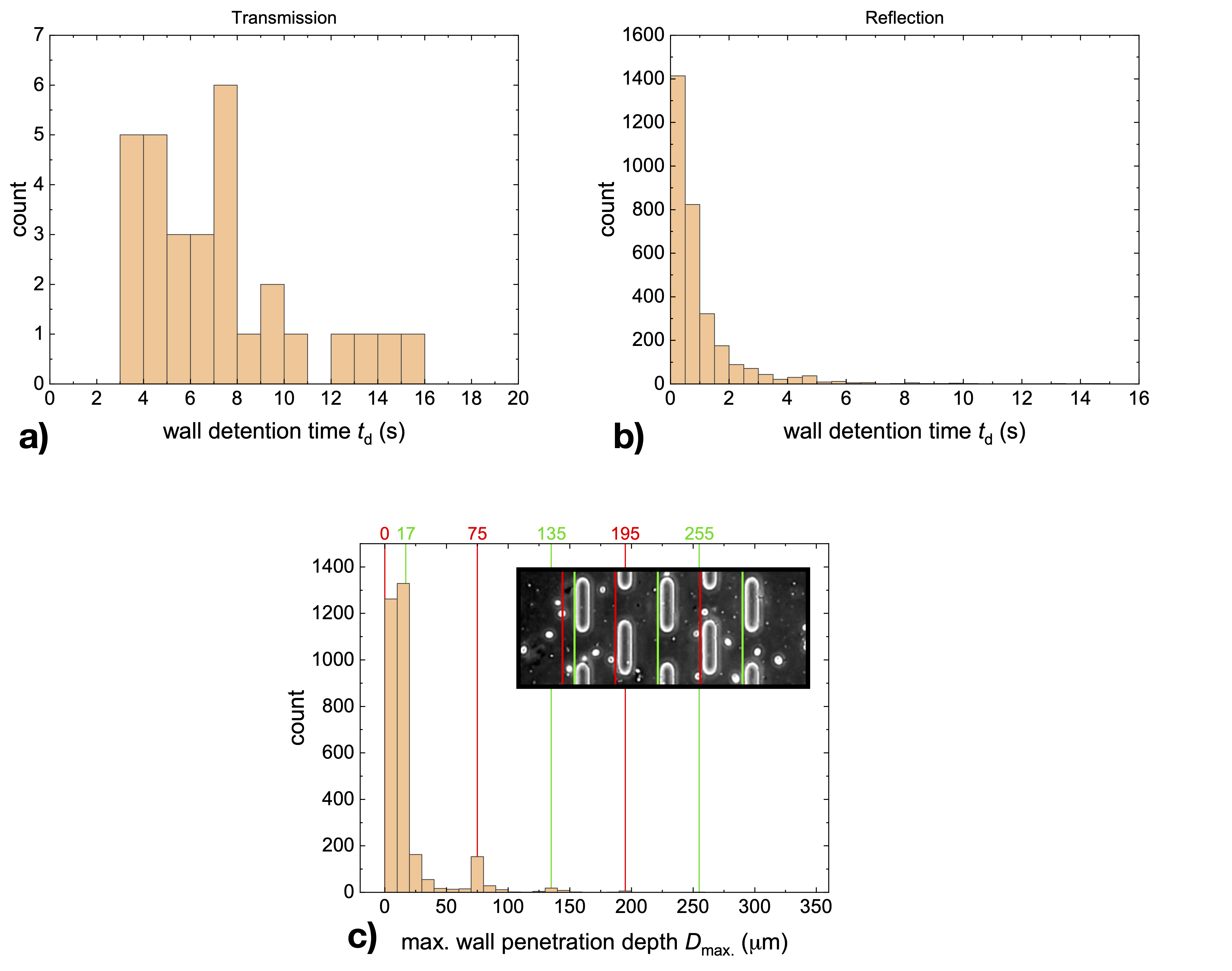}
\caption{Distribution of wall detention times for a) permeating and b) reflecting trajectories for Sample \textbf{C} with elongated pillars. c) Corresponding distribution of the maximal penetration depth. }
\label{fig:per_trajectories_stat}
\end{figure}

In the case of elongated pillars oriented perpendicular to the capillary axis (Sample \textbf{C}), the wall detention time for permeating algae, \(\tau^\text{perm}_\text{det} \approx\qty{7.2}{s}\),  becomes larger than in the case of parallel aligned pillars (Figure~\ref{fig:per_trajectories_stat}). This can be attributed to the trapping through the alignment of the trajectories in the orthogonal direction (see Figure \ref{fig:Trajectories}e,g). Indeed, the angular displacement distribution is strongly anisotropic, with maxima aligned orthogonally to the capillary axis (Figure \ref{fig:DisplacementElongated}a-c).

It should be noted that trajectories whose distance to the lateral microchannel walls is smaller than the steric interaction distance are excluded from the data set.
In the case of Sample \textbf{C},  this results in a small set of only 30 transmission trajectories which makes claims about the permeation detention time disputable.  Yet,  when letting go of the constraint and taking into account trajectories that may include steric wall interactions,  the wall detention time for permeation,  \(\tau^\text{perm}_\text{det} \approx\qty{6.2}{s}\) (from 283 trajectories,  see also Figure  S6),  is still outstandingly high.

The different behaviour of the motile algae, depending on the geometry of their surrounding,  is reflected by the distributions of the path length of  permeating trajectories. Two aspects must be considered when comparing path lengths: The availability of short paths and the probability of swimmers finding short paths. The former aspect is purely geometric, whereas the latter involves swimmer-obstacle interactions.  If we are interested in both aspects,  we contemplate the mean transmission path length \(\bar{L}_\text{trans}\) measured in units of the width \(w\) of the compartment \textbf{3} with obstacles.  
Alternatively, we consider the mean transmission path lengths normalized by the length of the shortest available path \(L^\text{min}_\text{trans}\) which reflects to what extent the geometry guides swimmers along short paths.  The quantity \(\bar{L}_\text{trans}/L^\text{min}_\text{trans}\) serves as a measure for the enhancement or hindrance of permeation by a microlabyrinth and emphasizes the effects of swimmer-obstacle interactions,  i.e.  alignment or randomization of swimmer motion.  
The shortest available paths are straight lines following obstacle-free lanes from compartment \textbf{1} to \textbf{2}.  In some cases,  algae might be able to find slightly smaller paths if they exploit the finite lane width to travel under a smaller angle to the horizontal.  Practically,  this deviation seems negligible.

The shortest paths can be found for samples \textbf{D} and \textbf{E} with parallel pillar orientation,  where a straight path parallel to the microchannel walls exists.  
These are the samples with the smallest mean transmission path lengths in units of \(w\) (see Table \ref{tab:Pathlength}). The geometry in Sample \textbf{D} also favours the shortest path,  as shown by the smallest value of \(L_\text{trans}/L^\text{min}_\text{trans}\) across all samples.  In contrast,  \(L_\text{trans}/L^\text{min}_\text{trans}\) in Sample \textbf{E} is larger than for the Samples \textbf{A} and \textbf{B} with cylindrical obstacles.
This implies that the parallel obstacles enhance permeation efficiency for the case with a larger porosity (Sample \textbf{D}) and reduce it for the smaller inter-obstacle distance in Sample \textbf{E}.  Swimmers are more likely to change lanes in \textbf{E} than in \textbf{D} (see Figure S4) which increases the path length.

Not surprisingly,  Sample \textbf{C} with perpendicularly oriented elongated obstacles has the longest mean transmission path length.  One should mention that,  due to the strong inhibition of transmission,  the values for \(\bar{L}_\text{trans}\) in Table \ref{tab:Pathlength} are based on only 30 data points.   Letting go of the restriction that only trajectories that stay far from the lateral microchannel walls are taken into account,  one obtains  values \(L_\text{trans}/w=2.36,\,L_\text{trans}/L^\text{min}_\text{trans}=1.67\),  showing a similar trend, from a much larger data set of 283 transmission events. The mean reflection path lengths of most samples are comparable,  with the exception of Sample \textbf{C} where a large fraction of the channel cross-section is occupied by obstacles,  resulting in frequent reflection at the first row of obstacles and consequently low \(\bar{L}_\text{refl}\).

\begin{table}[ht!]
\centering
\caption{Mean path lengths of trajectories in the compartment \textbf{3} with obstacles for reflected and transmitted algae.  Mean path lengths are normalized by the width of the region width of the compartment \(w\) or by the shortest possible path length for transmission \(L_\text{trans}^\text{min}\). }
\begin{tabular}{@{}llll@{} }
\hline
\text{Sample} & \(\bar{L}_\text{refl}/w \)& \(\bar{L}_\text{trans}/w\) & \(\bar{L}_\text{trans}/L_\text{trans}^\text{min}\)\\
\hline
A, \(\circ\) & 0.65 & 1.74 & 1.51  \\
B, \(\circ\) & 0.56 & 1.72 & 1.49 \\
C, \(\perp\) & 0.29 & 2.68 & 1.90   \\
D, \(\parallel\) & 0.46 & 1.37 & 1.37 \\
E, \(\parallel\) & 0.52 & 1.67 & 1.67 \\
\hline
\end{tabular}

\label{tab:Pathlength}
\end{table}

\subsection{Transmissivity and reflectivity}

Excursions of microswimmers through the compartment \textbf{3} with the obstacle array can be classified as transmission or reflection events to define reflection and transmission coefficients \(R\) and \(T\) as the ratio of the number of reflection/transmission events and the total number of observed excursions through compartment \textbf{3}.  The boundaries of compartment \textbf{3} are located \(\qty{17}{\mu m}\) from the outermost obstacle row (see also Figure \ref{fig:trajectories_cyl}), which represents the distance below which steric swimmer-obstacle interactions can occur (assuming a swimmer body radius of \(\qty{5}{\mu m}\) and a flagellar length of \(\qty{12}{\mu m}\) \cite{Jeanneret2016}).  
Transmission and reflection events occuring close to the lateral channel walls are discarded to exclude the effects of steric swimmer-wall interactions that may occlude the influence of obstacle geometry and arrangement.
\begin{table*}[ht!]
\centering
\caption{Transmission and reflection coefficients \(T\) and \(R\) for \chlamy at obstacle arrays of different porosity \(P\).  Obstacles were either cylindrical (\(\circ \)) or elongated in which case the orientation of their long axis was either parallel (\(\parallel\)) or perpendicular (\(\perp\)) to the main axis of the channel.  See Figure  \ref{fig:Samples} for details of the channel geometry for different samples. Porosities were determined from images.  }
\begin{tabular}{@{}llll@{} }
\hline
Sample & Porosity, \(P\) & Transmission, \(T\)  & Reflection, \(R\)\\
& (\%) & (\%) & (\%)\\
\hline
A, \(\circ\) & 94 & 30 & 70  \\
B, \(\circ\) & 97 & 32 & 68 \\
C, \(\perp\) & 78 & 1 & 99   \\
D, \(\parallel\) & 83 & 22& 78 \\
E, \(\parallel\) & 77 & 19 & 81 \\
\hline
\end{tabular}
\label{tab:CR:RT}
\end{table*}
Reflection and transmission coefficients for different channels are listed in Table \ref{tab:CR:RT}. 
The transmission was strongly suppressed when elongated obstacles were oriented perpendicularly to the channel's main axis, and nearly all algae were reflected (Sample \textbf{C},  Table  \ref{tab:CR:RT}).  Transmission coefficients for parallel orientation of elongated obstacles and cylindrical pillars were significantly larger than for perpendicularly oriented elongated obstacles.

Two effects contribute to the suppression of transmission for perpendicularly oriented stadium-shaped obstacles: Firstly,  this geometry has a low aperture,  i.e. the obstacles occupy a large part of the channel cross-section.  Secondly,  the preferred swimming direction of \chlamy inside the porous wall is unfavourable for transmission (Figure \ref{fig:DisplacementElongated}a-c).

Since perpendicular obstacles direct algae toward the lateral microchannel walls,  one might be concerned that the transmission/reflection statistics get distorted by the aforementioned exclusion of trajectories approaching the channel walls.  While the absolute number of transmission events increases substantially if the trajectories in the vicinity of the channel walls are considered,  the transmission coefficient for perpendicular obstacle orientation takes the value \(T=4\,\%\), thus remaining remarkably small.

A higher porosity leads to a larger transmission probability for channels with the same obstacle shape and orientation (compare Samples \textbf{A} and \textbf{B} or \textbf{D} and \textbf{E},  Table \ref{tab:CR:RT}).

\section{Conclusion}
In this paper, we demonstrated that the diffusion of the microswimmers \chlamy through a porous wall is strongly affected by the shape anisotropy of the obstacles comprising the porous medium.  Describing the transport using reflection and transmission coefficients, we show that the geometrical constraints and density of the obstacles determine the permeation through the wall and the reflection of the microswimmers. Reduction in the porosity of the wall results in the obstruction of the permeation of the microswimmers and enhancement of the reflection.  
  
Through the steric and hydrodynamic interactions between the microswimmers and the pillars, the shape anisotropy of the pillars results in the alignment of the microswimmers' trajectories introducing bias in the diffusive process. Scattering at the pillar arrays appears to contribute the most to the reflection events.
  
Depending on the pillars' orientation and mediated by the interactions with the microchannel's walls, such alignment may even lead to trapping, as was observed in the case of orthogonally oriented pillars  in Sample \textbf{C}.

\section{\label{sec:exp}Experimental}

\subsection{Cultivation of \chlamy}
\chlamy of the strain SAG 11-32a were purchased from the culture collection of algae at G\"{o}ttingen university.  The cells were grown under illumination by fluorescent lamps (Osram Fluora L15W/77) on a 14 h:10 h day-night-cycle in Tris-acetate-phosphate-medium (TAP-medium).  Air bubbling promoted gas exchange in the liquid cultures and,  by inducing flow,  mitigated biofilm formation. 

The strain was maintained by weekly subculturing.  On the 7\textsuperscript{th} day after the inoculation of a culture,  its cell density was determined using a hemocytometer.   Then,  a defined volume of the liquid culture was transferred to a new flask with TAP in a clean bench to inoculate a fresh culture with an initial cell density of \(10^5\) cells per ml.
Experiments were performed with vegetative cells taken from the cultures during the daytime on the 6\textsuperscript{th} to 8\textsuperscript{th} day after inoculation.

\subsection{Microfluidic chip fabrication}
Microfluidic labyrinths were fabricated by replicating the master wafer in the transparent, flexible polymer Polydimethylsiloxane (PDMS). The structures were designed and sketched using AutoCAD (Autodesk) software.  Next, the soft lithography mask based on Soda Lime (c) glass,  Al and Cu was fabricated at Compugraphics Jena GmbH according to the prepared sketch. 

The master mold fabrication was implemented using a soft lithography process on a 4'' silicon wafer with softened process parameters. Mainly, the temperature of the soft bake and post-exposure bake was reduced compared to a standard to ensure the ablation of the photoresist in circular openings with a radius of \qty{10}{\um}.  Thus, the silicon wafer was dehydrated at \qty{200}{\celsius} over 5 minutes, and the photoresist SU8-25 (MicroChem GmbH) was spin-coated (LabSpin 6, S\"{u}SS MicroTec GmbH) on it at 750 rpm for 6 seconds and subsequently at 2000 rpm for 30 seconds. The soft bake step was done at \qty{65}{\celsius} over 3 minutes, followed by \qty{85}{\celsius} over 10 min.  After the samples were completely dry, they were exposed through the prefabricated shadow mask with an exposure dose of \qty{160}{mJ/cm^2}  and wavelength of \qty{365}{\nm}. Post-exposure bake (PEB) was done at \qty{65}{\celsius} over 2 minutes, followed by \qty{85}{\celsius} over 10 min.  After PEB, the samples were slowly cooled down over 2 hours.  Finally, wafers were washed for 17 minutes with constant stirring in the mr-Dev600 developer to remove the non-exposed photoresist. After development, the master molds were immersed in Isopropanol for 2 minutes and dried with a nitrogen gun.

The microstructures of the master mold (Figure S1a) were pre-sputtered with 25 nm gold and analyzed using a profilometer (FRT MicroProf) and a scanning electron microscope (SEM, Zeiss EVO50). 

To obtain PDMS microfluidic channels, silicone-based polymer was mixed with 10:1 base to a curing agent, respectively, and poured on the prepared wafer. Curing took place at room temperature for over 48 hours. Once the PDMS was cured, it was removed from the mold and cut. Next, the \qty{170}{\um} thick glass slides and PDMS structures (Figure  S1b) were treated with oxygen plasma for 1 minute, brought into contact, and heated for over 10 minutes at \qty{80}{\celsius}.

Once the PDMS was bonded to the glass slide, the microfluidic chips were filled with DI water immediately to exploit the temporary hydrophilicity of the PDMS and, in this way,  avoid air bubbles in the microlabyrinth.

\subsection{Preparation of samples}
 \chlamy were extracted from culturing flasks under sterile conditions.  The suspensions of algae were centrifuged (\(1000\, g\) for \qty{10}{\minute}) and the supernatant was removed.  Subsequently,  the cells were re-dispersed in fresh TAP-medium.  

The sample was then filled into the PDMS-channels using a pipette.  The channel inlets were sealed with grease and the microfluidic chips were kept in water under the same lamp that was used for cell culturing for several hours.  It appears,  that a few hours after filling the channels,  cells are slow and sparse in the channel.  If channels are incubated overnight before usage,  more cells had spread from the inlets into the channel and the swimmers tended to be faster.

In some cases,  the light-switchable  adhesion of \chlamy \cite{Kreis2018} was exploited to prepare concentration gradients.  
Blue light was applied locally using a halogen lamp (Zeiss HAL 100) with a filter (\(\lambda\approx \qty{480}{nm},\, I\approx \qty{750}{\mu W /cm^2}\)).  
The cells were exposed to ambient red LED illumination (\(\lambda\approx \qty{628}{nm},\, I\approx \qty{110}{\mu W /cm^2}\)) to keep cells motile in the whole microchannel.
Swimming cells that reached the blue spot often switched into the adhered state.  The channels were kept under these conditions for about one hour to accumulate adhered cells in the blue spot.  Then,  the red ambient illumination was switched off and global blue illumination was applied for about 10 minutes using  a cold light source (Schott KL 2500 LCD) and a  blue bandpass filter (\(\lambda_\text{max}\approx\qty{513}{nm}, \, I\approx\qty{335}{\mu W /cm^2}\)).  
In this way,  cell adhesion was triggered globally.  Total adhesion was not achieved,  but a significant gradient of the adhered cell concentration could be prepared in this way.  The cells switched back into the swimming state after the illumination conditions were changed to red ambient light and red observation illumination.

\subsection{Observation}
An inverted microscope (Zeiss Axio Observer.D1) equipped with a digital camera (Canon EOS M6 Mark II) was used to record videos of \chlamy motion in the patterned microchannels (see Figure \ref{fig:Samples}a) at the resolution of 1920x1080 \(\rm{px}^2\) and the frame rate of \(\SI{50}{s^{-1}}\).

The microscope was used in transmission in phase contrast mode with a 10x-objective (Zeiss A-Plan 10x/0,25 Ph1).
The halogen lamp (Zeiss HAL 100) serving as a light source for observation was operated through a red glass filter (\(\lambda\geq \qty{630}{nm}\)) to enhance the cell motility \cite{Sineshchekov2000,Kreis2018}.  Additionally,  red ambient light was applied with LEDs (\(\lambda\approx \qty{628}{nm},\,I\approx 110\,\unit{\mu W/cm^2}\)).

\subsection{Cell tracking}

Videos of \chlamy in patterned microchannels were cropped to a size of 920x920 \(\rm{px}^2\) and rotated to achieve horizontal orientation of the microchannel borders.  To facilitate the detection of moving algae, the background was subtracted using a mixture of gaussian algorithms (BackgroundSubtractorMOG2 from OpenCV in Python) or minimum projection (in Matlab).  

The background image was refreshed as the video progressed (typically in intervals of 1000 frames) to account for dynamic changes due to the adhesion and detachment of algae at the glass bottom of the channel.

The detection and tracking was performed using Trackmate plugin in Fiji (ImageJ) \cite{Tinevez2017,Ershov2022} employing a Laplacian of Gaussians filter for edge detection and a Kalman tracker to extract the trajectories.

\medskip
\textbf{Supporting Information} \par 
Supporting Information is available from the Wiley Online Library or from the author.

\medskip
\textbf{Acknowledgements} \par 
The authors declare equal contributions to this work.
The authors thank Prof. Andreas Menzel and Dr. Dmitry Puzyrev for fruitful discussions and German Research Association (DFG) for funding (Project ER 467/14-1). FvR  acknowledges support from a Landesstipendium Sachsen-Anhalt. 

\medskip
\textbf{Conflict of Interest}\par
The authors declare no conflict of interest.

\medskip

\bibliographystyle{MSP}
\textbf{References}\\




\begin{figure}
\textbf{Table of Contents}\\
\medskip
  \includegraphics{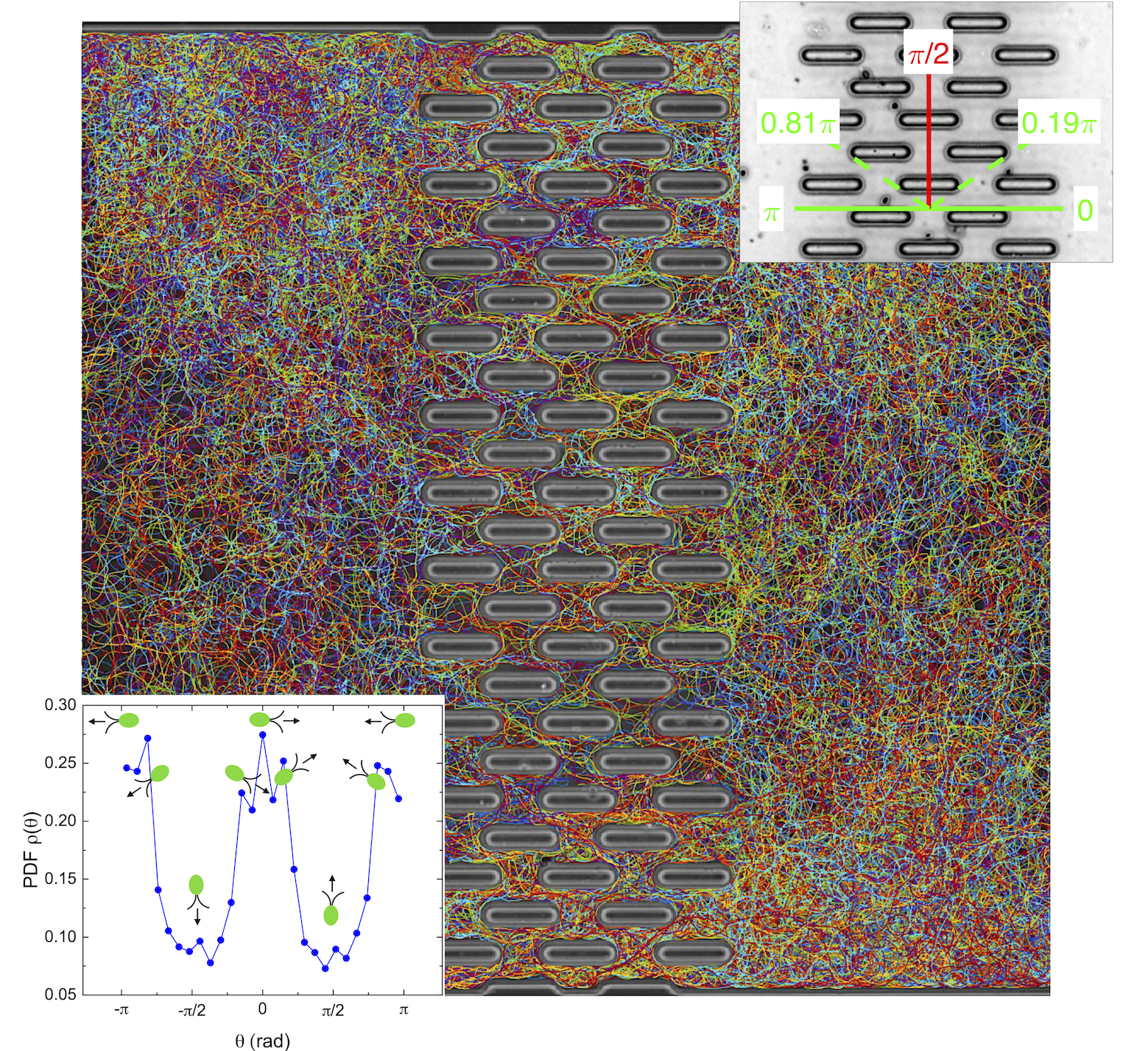}
  \medskip
  \caption*{Quasi two-dimensional PDMS-channels with arrays of pillars are used as a model system for a porous medium to study the effect of environmental anisotropy on the behaviour of motile microalgae. The geometry of the surrounding gives rise to preferred swimming directions and affects the permeation probability. }
\end{figure}

\end{document}